\documentclass[11pt]{JHEP3}
\usepackage{amsmath,epsfig}
\usepackage{amssymb,amsfonts}
\usepackage{graphicx}
\usepackage{multirow,cite}
\usepackage{subfigure}
\usepackage{datetime}
\newcommand{\bi}{\begin{itemize}}
\newcommand{\ei}{\end{itemize}}
\newcommand{\non}{\nonumber}
\def\half{\frac{1}{2}}
\def\p{\partial}
\def\a{\alpha}
\def\b{\beta}
\def\d{\delta}
\def\g{\gamma}
\def\l{\lambda}
\def\tl{\tilde{\lambda}}

\def\e{\epsilon}
\def\s{\sigma}
\def\r{\rightarrow}

\def\Om{\Omega}

\def\O{\mathcal{O}}
\def\R{\mathbb{R}}
\newcommand{\bea}{\begin{eqnarray}}
\newcommand{\eea}{\end{eqnarray}}
\newcommand{\be}{\begin{equation}}
\newcommand{\ee}{\end{equation}}

\title{Decrypting the warped black strings}
\author{Monica Guica\\

\vspace{1mm}

\hspace{-4.5mm}{\small   David Rittenhouse Laboratory, University of Pennsylvania, \\  \hspace{-0.55 cm}  Philadelphia, PA 19104-6396, USA}}

\abstract{

\bigskip

We propose a map for extracting the holographic meaning of the metric %and other background fields  
for  a class of warped AdS$_3$ black strings in type IIB supergravity.
 Our choice of holographic data is based upon a general prescription due to Papadimitriou. With this choice, we find a holographic stress tensor that is both symmetric and conserved, at least when restricted to a particular sector of the theory. Using the  holographic stress tensor, we compute the energy and momentum per unit length of the black strings, as well as   the  conformal anomaly,   and we show that the results agree with those previously obtained  via covariant methods.

 }

\begin{document}
%\maketitle

\section{Introduction}

Holography is believed to be a fundamental property of quantum gravity \cite{'tHooft:1993gx}; nevertheless, the holographic dictionary has only been understood in detail for the case of the AdS/CFT correspondence \cite{Maldacena:1997re,Aharony:1999ti,Witten:1998qj,Gubser:1998bc,deHaro:2000xn,Skenderis:2002wp} and % - usually to a lesser degree - 
a few other, qualitatively similar, cases. Examples include Lifshitz spacetimes \cite{Ross:2011gu,Mann:2011hg,Baggio:2011cp,Ross:2009ar,Korovin:2013bua}, non-conformal branes \cite{Wiseman:2008qa,Kanitscheider:2008kd}, non-AdS/QCD \cite{Aharony:2002up,Erdmenger:2007cm,Zaffaroni:2005ty,refsqcd}, etc.  

It is interesting to inquire whether there exist any tractable examples where the holographic dictionary is not just a straightforward extension of the AdS/CFT one. It appears that  Schr\"{o}dinger spacetimes \cite{Son:2008ye}, which geometrically realize the non-relativistic conformal group, provide such an example. Aside from their theoretical appeal, Schr\"{o}dinger spacetimes also have many potentially interesting applications, for example as gravity duals to strongly-coupled non-relativistic conformal field theories \cite{Son:2008ye,Balasubramanian:2008dm} and, in  the three-dimensional case\footnote{Three-dimensional Schr\"{o}dinger spacetimes are also known as null warped AdS$_3$. Their finite-temperature version, which is the subject of this paper, is called (spacelike) warped AdS$_3$.}, as toy models for  the Kerr/CFT correspondence \cite{kerrdip,kerrcft}.

The main challenge in understanding  holography for Schr\"{o}dinger spacetimes  is how to correctly identify the holographic sources to which boundary operators - and in particular the dual stress tensor - couple. In a beautiful paper \cite{papadimitriou}, Papadimitriou gave a general prescription for choosing the holographic data in an arbitrary spacetime, at least at the classical level. In \cite{fg} this prescription was applied to study %non-linear 
pure gauge modes in  three-dimensional Schr\"{o}dinger spacetimes and relate them to sources and expectation values for the putative holographic stress tensor. Although the dual theory is believed to be  non-local and non-relativistic \cite{Alishahiha:2003ru,maldanr,Herzog:2008wg,Adams:2008wt}, % (at least for this restricted set of modes) 
the holographic stress tensor   uncovered in \cite{fg} is both symmetric and conserved, signaling the presence of an emergent % emergence of a
relativistic symmetry.

Hints of a relativistic conformal symmetry \cite{Anninos:2008fx,Compere:2008cv} have long been present in the study of warped AdS$_3$ black hole/string backgrounds, which in the context of string theory correspond to finite-temperature states in three-dimensional 
Schr\"{o}dinger spacetimes. Asymptotic symmetry group analyses of these spacetimes reveal properties akin to those of two-dimensional \emph{relativistic} conformal field theories, such as the presence  of two  (mutually exclusive) Virasoro algebras and the applicability of Cardy's formula for the entropy \cite{stromwei,trunc}.

%That the dual field theory is secretly relativistic  - at least in the large $N$, strong coupling limit - has been long hinted at by the asymptotic symmetry group analyses in warped AdS$_3$ black hole/string backgrounds, which correspond to finite-temperature states in $3d$ %three-dimensional 
%Schr\"{o}dinger spacetimes. These analyses reveal properties akin to those of two-dimensional \emph{relativistic} conformal field theories, such as the presence  of two  (mutually exclusive) Virasoro algebras and the applicability of Cardy's formula for the entropy \cite{stromwei,trunc}.

%Warped AdS$_3$ black holes/strings - which correspond to finite-temperature states in three-dimensional Schr\"{o}dinger spacetimes - provide further evidence that the dual field theory shares certain properties with two-dimensional \emph{relativistic} conformal field theories. This evidence consists of  the presence  of two  (mutually exclusive) Virasoro asymptotic symmetry algebras and the Cardy asymptotic growth of states  \cite{stromwei,trunc}.

%have always been puzzling from a holographic point of view. On the one hand, one expects the dual field theory to be non-local and non-relativistic \cite{maldanr,Herzog:2008wg,Adams:2008wt}; on the other hand, asymptotic symmetry group analyses reveal properties akin to those of two-dimensional \emph{relativistic} conformal field theories, such as the presence  of two  (mutually exclusive) Virasoro algebras and the applicability of Cardy's formula for the entropy \cite{stromwei,trunc}.

In this note, we  propose an extension of the dictionary in \cite{fg} to a family of  warped AdS$_3$ black string solutions in string theory found in \cite{trunc}. As in the pure Schr\"{o}dinger case, the holographic stress tensor we find is both symmetric and conserved, in accordance with the results of the asymptotic symmetry group analyses. We also show that the expectation value of the  holographic stress tensor and the conformal anomaly in the warped black string backgrounds %and show that the values we obtain 
agree with the energy and momentum of the strings, as well as the conformal anomaly computed via other means in \cite{stromwei,trunc}. This match brings further support in favour of the holographic dictionary we are proposing.

Our dictionary, which is based on Papadimitriou's prescription, is not the only dictionary that has been put forth for understanding the holographic stress tensor in Schr\"{o}dinger spacetimes. Other attempts include \cite{Ross:2009ar,vanRees:2012cw}, of which  \cite{vanRees:2012cw} has carried out an extremely  careful and detailed analysis. These prescriptions are different from ours in that they choose different holographic sources, which results in different operator expectation values and different holographic Ward identities. While the tests that our proposal passes are certainly encouraging,
they only involve matching a gravity calculation to another gravity calculation. Ultimately, one should use input from the dual field theory  in order to decide on the correct holographic data
and the Ward identities they satisfy\footnote{
In certain cases, the dual field theory to Schr\"{o}dinger spacetimes is known: it is a  so-called null dipole theory \cite{Bergman:2000cw,Bergman:2001rw,Dasgupta:2001zu} or an  RG flow thereof. Dipole theories are similar to non-commutative ones, so it may be possible to find - perhaps using the methods of \cite{Okawa:2000sh,Liu:2001ps} -  the gauge-invariant operator that couples to the spacetime metric.}.

This paper is organised as follows. In section \ref{setup}, we %give the necessary background and 
review the holographic proposal of \cite{fg} for three-dimensional Schr\"{o}dinger spacetimes, as well as the simplest warped black string solution of \cite{trunc}. In section \ref{dipcft}, we discuss the modifications of this proposal that are necessary  for applying it  to the warped black string backgrounds. In section \ref{holo}, we construct a truncated holographic dictionary and compute the expectation value of the stress tensor for all the black string backgrounds of \cite{trunc}. We conclude with a discussion in section \ref{disc}.

\section{Setup and review \label{setup}}

We start from $AdS_3$ in Poincar\'e coordinates, with metric
\vspace{0.1cm}

\be
\frac{ds^2}{ \ell^2}= \frac{dw^+ dw^- + dz^2}{z^2}
\ee
This spacetime has $SL(2,\mathbb{R})_L \times SL(2,\R)_R$ isometry. The $SL(2,\R)_R$ generators are given by (minus) $H_i$, where

\be
H_{-1} = \p_+ \;, \;\;\;\;\; H_0 = w^+ \p_+ + \half z \p_z \;, \;\;\;\;\; H_1 = (w^+)^2 \p_+ + w^+ z \p_z - z^2 \p_-
\ee
and the same expressions, with $w^+\, \leftrightarrow \, w^-$, yield the $SL(2,\R)_L$ ones. Lowering the index on the above Killing vectors% (and multiplying by $-2/\ell^2$)
, we find  the left-invariant one-forms 

\be
\s_{-1} = \frac{dw^-}{ z^2} \;, \;\;\;\;\; \s_0 = \frac{w^+ dw^-}{ z^2} + \frac{dz}{ z} \;, \;\;\;\;\; \s_1 = (w^+)^2 \frac{dw^-}{ z^2} +2 w^+ \frac{dz}{z} - dw^+
\ee
These one-forms are all invariant under $SL(2,\mathbb{R})_L \times U(1)_R$, where for each of the forms $\s_i$,
the $U(1)_R$ is generated by the Killing vector $H_i$ and is non-compact. 

%the BTZ black holes in $AdS_3$ are obtained via the coordinate transformation

%\be
%w^\pm = \sqrt{\frac{r-T_+T_-}{r+T_+ T_-}} \, e^{2 T_\pm x^\pm} \;, \;\;\;\;\; z = \sqrt{\frac{2 T_+ T_-}{r+ T_+ T_-}} \, e^{T_+ x^+ + T_- x^-} \;, \;\;\;\;\; x^\pm = \phi \pm t
%\ee 
%followed by a quotient along $\phi$, $\phi \sim \phi + 2 \pi$. They correspond to finite-temperature states of the theory on a circle, with temperatures

%\be
%T_L = \pi^{-1} T_- \;, \;\;\;\;\; T_R = \pi^{-1} T_+
%\ee
%It is also possible to not identify the coordinate $\phi$. This still corresponds to a finite temperature state of the theory on a line, but the spacetime metric is still AdS. 

%\subsection*{Three-dimensional Schr\"{o}dinger spacetimes}

%\subsubsection*{Three-dimensional Schr\"{o}dinger spacetimes}

\bigskip

\noindent \emph{$\diamond$ Three-dimensional Schr\"{o}dinger spacetimes}

\medskip

\noindent A three-dimensional Schr\"{o}dinger spacetime is obtained by deforming $AdS_3$ by the one-form $\s_{-1}$ dual  to the null Killing vector field  $H_{-1} = \p_+ $

\be
\frac{ds^2}{\ell^2} = \frac{dw^+ dw^- + dz^2}{z^2} - \frac{\l^2 (dw^-)^2}{z^4}  \label{schr}
\ee
By construction, the above spacetime has $SL(2,\mathbb{R})_L\times U(1)_R$ isometry, where the  $U(1)_R$ corresponds to translations along $w^+$.  Schr\"{o}dinger spacetimes are not solutions of Einstein gravity alone, one needs to add matter to support them. In many cases, the matter can be modeled by a (topologically) massive vector field
 
\be
 A  = \lambda'  \ell \, \s_{-1} = \frac{ \l' \ell \,dw^-}{z^2}
\ee
The relationship between $\lambda$ and $ \lambda'$ is fixed by the equations of motion. Any scalars that may be present in the action must be constant  in a Schr\"{o}dinger background, due to the $SL(2,\mathbb{R})_L$ symmetry. 
 
\bigskip
\noindent{\emph{$\diamond$ Field theory duals}}

\medskip

\noindent In general, the field theory duals to three-dimensional Schr\"{o}dinger spacetimes are not known. Nevertheless,
%Three-dimensional Schr\"{o}dinger spacetimes are described holographically by a theory we will call a ``dipole CFT''\footnote{Definition of dipole CFT}. 
in \cite{nr} it was argued that, at least at large $N$ and strong coupling, they can be effectively described as a finite deformation of a two-dimensional CFT by a $(1,2)$ primary operator. This operator, although  irrelevant from the point of view of the usual conformal group, is however exactly marginal with respect to the $SL(2,\mathbb{R})_L$ conformal symmetry
\be
S = S_{CFT} + \l \int  d^2 w\, \mathcal{O}_{(1,2)} \label{def}
\ee
The above equation provides a formal definition of the  field theory duals to three-dimensional Schr\"{o}dinger spacetimes, in the given region of parameter space. 
%This operator deformation is supposed to provide an effective description of the field theory in the given region of parameter space. 
In the following,  we will denote the theories defined via  \eqref{def} as \emph{dipole CFTs}\footnote{This terminology was used in \cite{stromwei} to denote the two-dimensional field theory dual to a particular three-dimensional Schr\"{o}dinger spacetime in string theory, which is described by the low-energy limit of the dipole-deformed \cite{Bergman:2000cw} D1-D5 gauge theory. Since we expect that all  field theories dual to three-dimensional Schr\"{o}dinger spacetimes  share a certain number of universal characteristics, herein we will denote  \emph{all} of them % two-dimensional field theories dual to a $3d$ Schr\"{o}dinger spacetime 
as  dipole CFTs.}.

\bigskip
\noindent{\emph{$\diamond$ Linearized holographic data}}

\medskip

\noindent The first step in understanding any holographic dictionary \cite{Skenderis:2002wp} is to find the most general solution to the asymptotic equations of motion and isolate its  normalizable and non-normalizable components. %According to the usual prescription, non-normalizable modes should correspond to holographic sources for the dual operators, while normalizable modes corresponds to dual operator expectation values.
 %At least at linearized level about the background, all solutions to the equations of motion are known and one can try identifying the holographic sources for the boundary operators present in the truncation (and consequently the holographic vevs). Nevertheless, the various proposals in the literature, notably \cite{balt} and \cite{fg}, disagree on what the natural/ correct holographic sources should be. We review herein the proposal of \cite{fg}, which is inspired from the general ideas of \cite{papadimitriou} reviewed in the introduction.  
At linearized level, all perturbations of  three-dimensional Schr\"{o}dinger spacetimes have been classified \cite{nr,fg}. They consist of  pure gauge modes (that can be induced by a diffeomorphism), which we will call T-modes, and ``true'' propagating degrees of freedom, which we call X-modes. % The latter have  characteristic $w^+$-momentum-dependent radial fall-offs at small $z$. %Thus, in radial gauge ($\d g_{zi} =\d g_{zz} =0$)
Both types of modes are present in the  linearized solution for the metric and the massive vector

\be
\delta g_{ij} = \d g_{ij}^T + \d g_{ij}^X \;, \;\;\;\;\;\; \d A_\mu = \d A^T_\mu + \d A^X_\mu \label{xtmodes}
\ee
%Each type of solution contains both normalizable and non-normalizable modes.
%Unlike in AdS, here there is no immediately obvious way of choosing the holographic sources among the various functions of $w^\pm$ that parametrize the solutions, which is the reason for the various different proposals in the literature.
Unfortunately,  there is no obvious choice of holographic sources among the various functions of $w^\pm$ that parametrize the above solutions, leading to the different proposals in the literature.
In the following, we review a general prescription due to Papadimitriou \cite{papadimitriou} that can determine  the holographic sources in an arbitrary spacetime with a timelike boundary.

\bigskip
\noindent{\emph{$\diamond$ Papadimitriou's prescription}}

\medskip

\noindent The proposal of \cite{papadimitriou} is based on the observation that holographic renormalization can be viewed  as a canonical transformation that renders the variational principle at the asymptotic space-time boundary well-defined. The role of time is played by the radial coordinate $z$. This canonical transformation diagonalizes the map between the cotangent bundle of the phase space, parametrized by $(dq^I, dp_I)$, and that of the space of asymptotic solutions, parametrized by $(d\a^I,d\b_I)$. Thus, holographic renormalization typically corresponds to

\be
\left(\begin{array}{c} q^I \\ p_I \end{array}\right) \underset{transformation}{\overset{canonical}{\xrightarrow{\hspace*{24mm}}}} \left(\begin{array}{c}\tilde q^I \\ \tilde p_I\end{array}\right) \;\;\sim\;\; \left(\begin{array}{cc} z^{-2\nu} & 0 \\ 0 & z^{2\nu} \end{array}\right)\left(\begin{array}{c}\a^I \\ \b_I \end{array}\right)
\ee
for some $\nu >0$.
%where $( q^I, p_I)$ parametrize the phase space and $(a^I, b_I)$, the space of asymptotic solutions. 
Since this  transformation is canonical, the %radial 
symplectic form is unchanged

\be
\Omega = d q^I \wedge d p_I =  d \tilde q^I \wedge d \tilde p_I = d \a^I \wedge d \b_I
\ee
In this framework, finding the holographic sources and expectation values is straightforward: one simply needs to evaluate the symplectic form on a general solution to the asymptotic equations of motion and then diagonalize it. The resulting diagonal entries are the holographic sources and expectation values.

It is not hard to show \cite{fg} that 
%Applying the above prescription to linearized perturbations of $3d$ Schr\"{o}dinger spacetimes, we find that for a general theory of gravity coupled to massive vectors and scalars, 
the symplectic form for  linearized modes in three-dimensional Schr\"{o}dinger spacetimes takes the diagonal form

\be
\Omega \sim \int T \wedge T + X \wedge X
\ee    
and moreover the source for the T-modes is a two-dimensional symmetric tensor $\d g^{(0)}_{ij}$. Applying the prescription above,
%Given that the symplectic form  does not have any cross terms between the X and T modes, 
we conclude that the X and the T modes couple to different  operators in the dual field theory, and in particular  the T modes  source the dual stress tensor.%, whereas the X modes  couple to the components of the massive vector. Note that, due to \eqref{xtmodes}, isolating either the T-modes or the X-modes in terms of the induced fields at the boundary, $h_{ij}$ and $A_\mu$, is extremely complicated. 

\bigskip
\noindent{\emph{$\diamond$ A Fefferman-Graham-like expansion}}

\medskip

\noindent If one is interested solely in the holographic stress tensor, it is possible to consistently set the X-modes to zero. In this case, it was shown in \cite{fg} that the general non-linear solution for the T-modes is given by
\be
g^T_{\mu\nu} = \hat g_{\mu\nu} - A_\mu^T A_\nu^T
\ee
where $\hat g_{\mu\nu}$ is the metric of an auxiliary $AdS_3$ spacetime of radius $\ell$ and takes the general Fefferman-Graham form. The vector field $A_\mu^T$ solves

\be
F_{\mu\nu} = \frac{2}{\ell}\, \hat \e_{\mu\nu\rho} A^\rho \;, \;\;\;\;\; A_\mu \hat A^\mu =0
\ee
The symplectic form evaluated on this solution is identical\footnote{Provided we use radial gauge in the Schr\"{o}dinger spacetime. Explicit linearized calculations show that the symplectic form is not conserved with respect to the radial coordinate of the auxiliary AdS$_3$, but it is conserved with respect to the Schr\"{o}dinger radial coordinate. %\emph{Check!} 
} 
to that  of the auxiliary AdS$_3$ spacetime. By Papadimitriou's prescription, we choose the holographic sources to be those of the auxiliary AdS$_3$. This choice leads to a holographic 
 %which prompts us to choose the holographic sources to be those of the auxiliary AdS. With this choice, 
%as is the holographic stress tensor. In particular, the 
stress tensor that  is symmetric, conserved and yields the same conformal anomaly as the $AdS_3$ one, at least when evaluated on the T-modes only.

% These properties suggest the presence of an enhanced relativistic conformal symmetry, at least when the X-modes are absent. 

\bigskip
\noindent{\emph{$\diamond$ The warped black strings}}

\medskip

\noindent The outcome of the analysis of \cite{fg} is rather intriguing and it would be useful to test its predictions in cases where an independent computation is possible. Black holes/strings provide a good testing ground, as one can use covariant methods \cite{Barnich:2001jy} to compute various non-trivial conserved charges and anomalies, which can then be compared with predictions from the holographic stress tensor.

%Nevertheless, warped analogues of the BTZ black hole (carrying both mass and angular momentum and non-extremal) do not easily exist in Schr\"{o}dinger spacetimes. Nevertheless, it is possible to construct families of them in type IIB string theory truncations \cite{trunc}.

The simplest theories of gravity coupled to massive vector fields do not seem to contain generic enough (non-extremal) black hole solutions. Nevertheless, such solutions do abound in string theory. 
In \cite{trunc}, a four-parameter family of stringy warped analogues of the non-extremal BTZ black string (carrying both mass and  momentum) 
has been constructed. % in type IIB string theory.
%While all these black holes share the same universal properties, here we will concentrate on the simplest example (the S-dual dipole black hole), while leaving the remaining solutions for section 6.
 The simplest warped black string in this family can be viewed as a solution of the following  three-dimensional action
\be
S= \frac{1}{16\pi G_3} \int d^3 x \sqrt{g} \left( R - 4 (\p U)^2 - \frac{4}{\ell^2} \, e^{-4U} \,  A^2 + \frac{2}{\ell^2} \, e^{-4U} (2-e^{-4U}) -\frac{1}{\ell} \, \e^{ijk} A_i F_{jk}\right) \label{actdip}
\ee
which descends from a consistent truncation of type IIB supergravity \cite{trunc}. The warped black string solution is
\be
\frac{ds_3^2}{\ell^2}  = T_+^2 dx_+^2  + \left[ T_-^2 (1+ \tl^2 T_+^2) - \tl^2 r^2\right] dx_-^2 + 2 r \, dx^+ \, dx^- + \frac{\ell^2  dr^2}{4 (r^2 - T_+^2 T_-^2)} (1+ \tl^2 T_+^2) \non
\ee

\be
A= \frac{\tl \ell}{\sqrt{1+\tl^2 T_+^2}} (r dx^- + T_+^2 dx^+) \;, \;\;\;\;\; e^{4U} = 1 + \tl^2 T_+^2 \label{bhmet}
\ee
The parameters $T_\pm$ are related to the Hawking temperature and horizon velocity  via

\be
T_H = \frac{2}{\pi} \frac{T_+ T_-}{T_+ + T_-} \;, \;\;\;\;\; \Om_H = \frac{T_+ - T_-}{T_+ + T_-} \label{tempom}
\ee
In \cite{trunc} it was shown that the energy and momentum per unit length of these black strings are given by 
\be
E \pm P = \frac{ \ell}{4 \pi G_3} T_\pm^2 \label{epmp}
\ee
Moreover,  asymptotic symmetry group analyses indicate the presence of either a left-moving or a right-moving Virasoro symmetry with central charge

\be
c  = \frac{3 \ell}{2 G_3} \label{cc}
\ee
These properties can be shown to hold with minor modifications  also for the general four-parameter family of warped black strings \cite{trunc}. We will come back to these more general solutions in section \ref{genbsb}.

\bigskip
\noindent{\emph{$\diamond$ Objectives}}

\medskip

\noindent We would like to  extract from the metric \eqref{bhmet} which operator sources and expectation values have been turned on, and to reproduce  $E,P$ and $c$ above from a holographic stress tensor computation, as has been done in \cite{Balasubramanian:1999re,Hollands:2005wt,Papadimitriou:2005ii} for the case of AdS. The challenge is that the black hole spacetimes are \emph{not} diffeomorphic to the vacuum Schr\"{o}dinger background, so the T-mode analysis in \cite{fg} does not strictly apply. In this paper we propose a generalization of \cite{fg} that reproduces the above quantities from a holographic computation.

\section{Dipole CFTs at finite temperature \label{dipcft}}

Any two-dimensional theory can be put at finite temperature\footnote{We will be interested in the more general setup where we turn on both a temperature $T_H$ and a  potential conjugate to momentum  $\Omega_H$, whose relationships to $T_\pm$  are given in \eqref{tempom}. The parameters $T_\pm$ are related to the usual left/right moving temperatures defined in e.g. \cite{strexcl} via $T_\pm = \pi T_{R/L}$.} by restricting the space where it is defined to a Rindler wedge of Minkowski space. The Rindler coordinates  $x^{\pm}$  are related to the Minkowski coordinates  $w^{\pm}$
via the coordinate transformation $w^\pm = e^{2T_\pm x^\pm}$.  In a two-dimensional conformal field theory, this is a conformal transformation which induces a thermal expectation value for the stress tensor.

A natural way to extend the definition \eqref{def} of dipole CFTs to Rindler space is

\be
S = S_{CFT} + \l \int  d x^+ dx^-\, \mathcal{O}_{(1,2)} (x^+,x^-) \label{deftemp}
\ee
%Note that the deforming operator is not invariant under conformal transformations, but the above 
which is the only $(1,2)$ deformation that is consistent with translation invariance in the $x^{\pm}$ plane.  We are assuming that $\l$ is $T_+$-independent. If we map the deformation  back to Minkowski space, we find

\be
\lambda \int d x^+ dx^- \mathcal{O}_{(1,2)} (x^+,x^-)\;\; \r \;\; 2 \lambda T_+ \int dw^+ dw^- w^+ \O_{(1,2)}(w^+, w^-) \label{plop}
\ee
where we  used the fact that $\O_{(1,2)}$ is a conformal primary. % of the indicated weights. 
The deformation on the right hand side still preserves $SL(2,\mathbb{R})_L \times U(1)_R$ isometry, but this time the $U(1)_R$ corresponds to scaling transformations instead of translations. We will assume that the $w^+$-dependent deformation above is  exactly marginal with respect to the left-moving conformal symmetry. This assumption  could be  verified using conformal perturbation theory, as in \cite{nr}.

%\emph{Can $\lambda T_+$ be corrected at higher orders in perturbation theory?} 

%The above deformation on the $x^{\pm}$ plane is the only translationally-invariant deformation that involves the $(1,2)$ deformation. Also, it gives a more-or-less sensible holographic dictionary. We could also consider the operator fixed on the plane + coordinate transformation. This does not correspond to the black hole, but to a particular time-dependent background. From the point of view of the $x^\pm$ observer, we are inserting some very funny, exponentially temperature-dependent operator.

The gravity dual of the deformation on the right-hand side of \eqref{plop} is naturally constructed in terms of the $w^+$-dependent $SL(2,\mathbb{R})_L \times U(1)_R$-invariant one-form on $AdS_3$,  $\s_0$,
\be
 \s_0 = \frac{w^+ dw^-}{ z^2} + \frac{dz}{ z} 
\ee
where the $U(1)_R$ represents right-moving scaling transformations. Because $\s_0 $ is not null, the background dual to the deformation will no longer be exact to second order in perturbation theory, as was true of the Schr\"{o}dinger spacetimes \cite{Kraus:2011pf}. Nevertheless, we can construct the solution order by order in perturbation theory\footnote{The identification of the coefficient of the boundary operator with the asymptotic value of the spacetime field, to all orders in perturbation theory, is based on the dictionary proposed in \cite{vanRees:2012cw,vanRees:2011fr,vanRees:2011ir}. Note that the resulting gravity solution makes sense only if $2\l T_+ < 1$. We are not sure whether this is a physical restriction on the theory - namely, that the inverse temperature cannot be smaller than the non-locality scale $\l$ - or the assumption that $\l$ is $T_+$-independent is wrong, or the all-orders holographic dictionary that we used is incorrect. In any case, as long as $\tilde \l = \l f(\l T_+)$ with $f(
0)=2$, the precise form of this holographic  map does not affect 
the computations of section \ref{holo}.} in $\lambda T_+$

\be
ds^2_{\s_0}= \frac{1}{1-4\l^2 T_+^2} \left( ds^2_{AdS_3} - A_{\s_0} \otimes A_{\s_0} \right) \;, \;\;\;\;\; A_{\s_0} = 2 \l  T_+ \ell \, \s_0 % \;, \;\;\;\;\; \mu = \frac{\kappa}{\sqrt{1-\kappa^2}} = \tl T_+
%\;, \;\;\;\;\; \mu = \tl T_+
\label{bhma}
\ee
The relationship between the parameter $\tl$ introduced previously in \eqref{bhmet} and the ``true'' coupling constant $\l$ of the dipole CFT is
\be
\tl = \frac{2\l}{\sqrt{1-4\l^2 T_+^2}}
\ee
%In the above, $\l$ is supposed to represent the coefficient of the deforming operator. The given relationship between $\tl$ and $\l$ is correct if the coefficient of the operator equals the boundary value of $A$. The relationship between $\tl$ and $\lambda$ is correct if there are no further corrections on the plane to the coefficient of the integrated operator. 
Performing the coordinate transformation

\be
w^\pm = \sqrt{\frac{r-T_+T_-}{r+T_+ T_-}} \, e^{2 T_\pm x^\pm} \;, \;\;\;\;\; z = \sqrt{\frac{2 T_+ T_-}{r+ T_+ T_-}} \, e^{T_+ x^+ + T_- x^-} \label{cootransf}
\ee 
we get precisely the warped black string background \eqref{bhmet}. This is the same coordinate transformation that takes Poincar\'e AdS$_3$ into BTZ \cite{strexcl}. Note that, asymptotically,  $w_\pm \sim e^{2 T_\pm x^\pm}$, as expected.

To summarize, our assumption \eqref{deftemp} implies %of the finite-temperature dipole CFT asserts 
 that from the point of view of the theory on the $x^\pm$ plane, the only operator source turned on is that of the defining operator. Each of the $x^\pm$ observers sees a thermal bath of particles  and a non-zero expectation value of the stress tensor. Meanwhile,  the Minkowski $w^\pm$ observer sees a non-trivial source for a $w^+$-dependent operator, whose coefficient depends on $T_+$. Thus, different thermal states in the dipole CFT, from the point of view of the Minkowski observer,  correspond to different $SL(2,\mathbb{R})_L$-invariant irrelevant deformations of a CFT$_2$ or, equivalently, to different exactly marginal deformations of a dipole CFT. Finally, from the bulk point of view, each  $w^+$-dependent irrelevant deformation corresponds to a different non-normalizable X-mode, with different asymptotics. Any given black string background can be obtained by turning on a $T_+$-dependent non-normalizable (but nevertheless tractable) deformation of AdS$_3$,
%, one needs to first turn on the correct $w^+$-dependent  X-mode, 
followed by the coordinate transformation \eqref{cootransf}, which  corresponds to a non-linear T-mode.

% corresponding to an X-mode, and then excited certain T-modes on top of this background.  %Nevertheless, the observer in the $x^\pm$ coordinates sees the $w^\pm$ vacuum as a thermal bath of particles. While all $x^{\pm}$ coordinates cover the same quarter of the $w^\pm$ Minkovski space, they are different for each value of $T_\pm$. 
%
%The unusual feature that we find is that the coefficient of the operator on the $w^+$ plane  depends on $T_+$, so different thermal states in the dipole CFT, from the point of view of the Minkovski observer,  correspond to different irrelevant deformations of AdS$_3$, with different asymptotics.

\section{Holography \label{holo}}
 
In the following, we would like to compute the expectation value of the holographic stress tensor in the warped black string backgrounds using holographic renormalization. Since from the point of view of each of the $x^\pm$ observers, only the stress tensor is turned on, we can compute its expectation value  by simply coupling the theory to a background metric and then varying  the on-shell action with respect to it. According to Papadimitriou's prescription, studying exclusively the  stress tensor sector corresponds to solely turning on T-modes around a given black string spacetime.

 Because backgrounds with different $T_+$ are not diffeomorphic to each other, we will not be able to change the state of the theory by just turning on T-modes. Thus, we will be forced to work at fixed $T_+$. Nevertheless, for every fixed $T_+$, we will  still be able to couple the theory  to an arbitrary metric (at least in the classical analysis) and compute the expectation value of the stress tensor. 
%This is because backgrounds with different $T_+$ are not diffeomorphic to each other, so we cannot study the relationship among them by just turning on T-modes.
%That is, for each of the $x^\pm$ observers, it is sufficient to just turn on a source for the T-modes. To change from one Rindler observer to another one needs to also modify the X-modes. Nevertheless, the fact that in each of these backgrounds T-modes correspond to stress tensor modes is simply a consequence of Papadimitriou's prescription; a similar prescription would be true for e.g. different 4d asymptotically flat black holes.
We  find  that the holographic dictionary - and in particular the designation of the holographic sources - depends on $T_+$, and thus on the state of the theory. This is  a joint consequence of  Papadimitriou's prescription  - which  identifies T-modes around any given background with the stress tensor sector - and the unusual feature of Schr\"{o}dinger spacetimes  that different states correspond to different asymptotic behaviors, as is clear from the temperature-dependence of the coefficient of the deforming operator %\footnote{The only way that the said coefficient will not be temperature-dependent is to have $\l =  const./T_+$, which has a singular zero-temperature limit. } 
on the right-hand side of \eqref{plop}.

\subsection{The finite-temperature FG expansion}

For reasons of computational simplicity, as in \cite{fg}, we restrict our study to T-modes only.
To construct general T-modes above the black string backgrounds, we first note that
the original black string metric \eqref{bhmet} can be written as\footnote{The coefficient of the $A_\mu A_\nu$ term is fixed by the requirement that certain divergent terms cancel between $g_{\mu\nu}$ and $A_\mu A_\nu$. The coefficient of $\hat g_{\mu\nu}$ is fixed by demanding that the symplectic form  not have any temperature-dependent overall factors. }

\be
g_{\mu\nu} = (1+\mu^2) (\hat g_{\mu\nu} - A_\mu A_\nu)\;, \;\;\;\;\;\;\; \mu \equiv \tl T_+ \label{defhatg}
\ee
where $\hat g_{\mu\nu}$ is the BTZ metric, satisfying

\be
\hat R_{\mu\nu} + \frac{2}{\ell^2} \, \hat g_{\mu\nu} =0 \label{eqadsmet}
\ee
and the background vector field $A_\mu$ solves

\be
F_{\mu\nu} = \frac{2}{\ell} \hat \e_{\mu\nu\l} A^\l \;, \;\;\;\;\; \hat A^2 = \frac{\mu^2}{1+\mu^2} \label{eqA}
\ee
Since both the definition \eqref{defhatg} and the equations of motion for $\hat g, A$ are covariant, it follows that the above equations hold for any background that is diffeomorphic to the original one. Thus, to find the most general T-mode solution around a given black string background one simply needs to solve \eqref{eqadsmet} for $\hat g$, then solve \eqref{eqA} for $A$ and plug into \eqref{defhatg}. The constraint $\hat A^2=const.$ guarantees (at least at linearized order) that only pure gauge modes are solutions. 

Writing the metric $\hat g$ in radial gauge

\be
d\hat s^2 =  d \eta^2  + \hat \g_{ij} dx^i dx^j  \label{radmet}
\ee
the general solution for $\hat \g$ is given by the Fefferman-Graham expansion \cite{Skenderis:1999nb}

\be
 \hat \g_{ij} = e^{2  \eta/\ell} \hat g^{(0)}_{ij} +  \hat g^{(2)}_{ij} +  \frac{1}{4} \, e^{-2 \eta/\ell} \hat g^{(2)}_{i}{}^k  \hat g^{(2)}_{kj}
\ee
where all indices are raised and lowered using the boundary metric $\hat g^{(0)}$.  The asymptotic equations of motion imply that

\be
\hat \nabla_i \hat g^{(2)i}{}_j = \hat \nabla_j \hat g^{(2)k}{}_k \;, \;\;\;\;\; \hat g^{(2)k}{}_k = - \half \hat R[\hat g^{(0)}] \label{constrg}
\ee
Note that if $\hat g^{(2)}_{ij}$ has a generic dependence on $x^i$, then it  is entirely  determined by $\hat g^{(0)}_{ij}$, albeit in a very non-local fashion. If the solution considered is constant in $x^i$, then the trace of $\hat g^{(2)}$ needs to vanish, but the remaining components are independent of $\hat g^{(0)}$. 

One may notice that \eqref{eqA} represent four equations for the field $A_\mu$, which only has three components. While the system seems overdetermined, a solution is always guaranteed to exist, because all solutions can be induced by a diffeomorphism. Nevertheless, one cannot solve for $A_\mu$ for an arbitrary background metric $\hat g_{\mu\nu}$, one needs to put the latter at least partially on-shell. More precisely, one needs to use multiples of \eqref{constrg}
in order to satisfy all equations $A$ obeys, but never needs to solve for the the non-local relationship between $\hat g^{(0)}$ and $\hat g^{(2)}$. In particular, for the case of constant sources, the full solution for both $\hat g$ and $A$ is parametrized by five independent constants. %\footnote{In the linearized approximation, all equations solved by $A_\mu$ become algebraic in $\rho$, not differential.}. 
%Note that we need to solve four equations for the three components  of $A$. Thus, the system seems over-determined. Nevertheless, a solution always exists, provided some of the equations of motion for $\hat g$ are used. The equations we need to solve for $A$ do not impose any non-local relationship between $\hat g^{(0)}$ and $\hat g^{(2)}$ in the asymptotic metric expansion, as we could deduce by studying the linearized model.
For non-constant sources, the leading term in the $A_\mu$ expansion bears a rather non-local relation to the boundary data $\hat g^{(0)}$.

Given the radial AdS$_3$ metric \eqref{radmet}, one finds that the corresponding warped AdS$_3$ metric reads\footnote{For future reference, we also have
\be
\g^{ij} = \frac{1}{1+\mu^2} \ \hat \g^{ij} + \frac{\hat A^i \hat A^j}{1+ (1+\mu^2) A_\eta^2} \;, \;\;\;\;\;\;\;\; \det \g = (1+\mu^2) (1+ (1+\mu^2) A_\eta^2) \det \hat \g
\ee}

\be
ds^2 = N^2 d \eta^2 + \g_{ij} (d x^i + N^i d \eta) (d x^j + N^j d  \eta)
\ee
with
\be
\g_{ij} = (1+ \mu^2)( \hat \g_{ij} -  A_i A_j )\;, \;\;\;\;\; N_i = - (1+ \mu^2) A_i A_\eta \;, \;\;\;\;\; N^2 = \frac{1+\mu^2}{1+ (1+\mu^2) A_\eta^2}
\ee
Unless $A_\eta =0$, if the AdS$_3$ metric is in radial gauge, then the warped AdS$_3$ one will not be, and vice-versa. Since in holographic renormalization it is natural to have the boundary at fixed radial coordinate, we need to bring the warped AdS$_3$ metric to radial gauge via a coordinate transformation. At linearized level above the black hole background, it reads

\be
 x^i \r x^i + \xi^i (\eta, x^i) \;, \;\;\;\;\; \;\; \xi^i = (1+\mu^2) \int d\eta \, A^i A_\eta \label{eqxi}
\ee
%we find, to first order in $A_\eta$
%\be d\tilde s^2 = (1+\mu^2)\, d\eta^2 + (\g_{ij} + \p_i \xi_j + \p_j \xi_i)\, dx^i dx^j   \;, \;\;\;\;\; 
%\ee
%
%\be \tilde A_i = A_i + \xi^k F_{ki} + \p_i (\xi^k A_k) \;, \;\;\;\;\; \tilde A_\eta = A_\eta 
%\ee
In the explicit linearized solution, $A_\eta$ is $\eta$-independent, as is the zeroth order $A^i$. Consequently, one can integrate  \eqref{eqxi} to find

\be
\xi^i =  \eta\, N^2 A^i  A_\eta \label{formxi}
\ee
%The issue of the correct radial coordinate is important because one can show that the symplectic form is conserved\footnote{Explicit linearized calculations show that the symplectic form is not conserved with respect to the auxiliary AdS radial coordinate $\rho$, but it is conserved with respect to $\rho'$.}. 
 %with respect to $\eta$, but not $\hat \eta$. Clearly, this issue does not arise in the constant case, where the two coordinates are just proportional.
One can then show that the radial symplectic form, when evaluated the general linearized T-mode solution%\footnote{Strictly speaking, we have only shown this for $x^i$-independent linearized perturbations around the black string metric. For general perturbations, the symplectic form is zero upon using the asymptotic equations of motion, as it should be in AdS$_3$.   }
, equals the radial symplectic form of the auxiliary AdS$_3$ spacetime\footnote{To show this, one needs to  use  various multiples of the linearized asymptotic equations of motion \eqref{constrg}. The calculation is almost identical to that in section $2$ of \cite{fg}. }. By Papadimitr‭iou's prescription, we choose the holographic sources to be $\hat g^{(0)}_{ij}$.

\subsection{Holographic renormalization}

%Note that the $AdS_3$ asymptotic expansion is performed in radial gauge in AdS. In general, the component $A_\rho$ of the vector field that satisfies \eqref{eqA} is non-zero, which implies that the warped AdS metric $g_{\mu\nu}$ \emph{is not} in radial gauge. Thus, we need to perform first a gauge transformation on $g_{\mu\nu}$, and put it in radial gauge with a new coordinate $\rho'$, then compute the regulated action with a UV cutoff at $\rho'=\e$ and renormalize it. This problem does not appear for the case of constant deformations, because $A_\rho=0$. Writing the $AdS_3$ metric in radial gauge

To perform holographic renormalization, we need to compute the on-shell action for the general T-modes, regulate it and then remove the divergences. The on-shell action consists of a bulk piece, which 
%
%\be S_{on-shell} =  S_{bulk} + \frac{1}{8\pi G_3} \int d^2 x  K \sqrt{\g} 
%\ee
%where $S_{bulk}$ 
is given by \eqref{actdip} evaluated on the general T-mode solution

\be
S_{bulk}= - \frac{1}{16\pi G_3}  \int d^3 x  \frac{4}{\ell^2 (1+\mu^2)} \sqrt{g} =  - \frac{1}{16\pi G_3}  \int d^3 x  \frac{4}{\ell^2} \sqrt{\hat g}  \label{sbulk}
\ee
and a Gibbons-Hawking boundary term\footnote{ Since ultimately we would like to keep the metric $\hat g$ fixed at the boundary rather than $g$, it is not clear whether the Gibbons-Hawking boundary term above is the one required by the modified variational principle. We decided to add it,  nevertheless, because we would like to view
 holographic renormalization as  a canonical transformation between the original variables $(\g_{ij}, \pi^{ij})$
and the new ones  $(\hat \g_{ij}, \hat \pi^{ij})$, and we require  the starting action to have 
a well-defined variational principle. }

\be
S_{GH} =  \frac{1}{8\pi G_3} \int d^2 x  \, K \sqrt{\g} 
\ee
Let us ignore for a moment the final diffeomorphism needed to bring the warped AdS$_3$ metric into radial gauge. In this case, \eqref{sbulk} shows that the bulk on-shell action is identical to that of the auxiliary AdS$_3$ spacetime. One can also easily compute the extrinsic curvature $K$ from its definition

\be
K_{ij} = - \frac{1}{2 N} (\p_\eta \g_{ij} - D_i N_j - D_j N_i)
\ee
where $D_i$ is the covariant derivative associated to the metric $\g_{ij}$. Plugging in the expressions for $N, N_i$, we find 
\be
 K_{ij} = \frac{ (1+ \mu^2) }{N} \hat K_{ij} - \frac{N}{2} (1+\mu^2)   (\hat D_i A_j + \hat D_j A_i)  A_\eta + \frac{N}{\ell} (A_i \hat \e_j{}^k + A_j \hat \e_i{}^k) A_k
\ee 
Taking the trace, one can show that

%\be 2 N  K = \frac{2 N^2 \hat K}{1+\mu^2} - \frac{2 (1+\mu^2)^2\hat A^i \p_i A_\eta}{(1+ (1+\mu^2)^2 A_\eta^2)^2}\ee
%Consequently,

\be
K \sqrt{\g} = \hat K \sqrt{\hat \g} - N^2  \hat A^i \p_i A_\eta \sqrt{\hat \g} \label{relK}
\ee
at full nonlinear level. Thus, barring the issues involving the final diffeomorphism, the on-shell action for T-modes in a warped black string background is identical to that of the auxiliary AdS$_3$ spacetime used to construct the T-modes, up to the %addition of a counterterm equal to the 
last term in \eqref{relK}. To render the action finite, we simply need to subtract this term  and
include an additional counterterm proportional to $\int \sqrt{\hat \g}$, as in AdS$_3$. Because the equations of motion for $A$ - which we used extensively in manipulating the above expressions -  do not impose any non-local constraint between the asymptotic metric coefficients, the resulting renormalized on-shell action will be the same functional of $\hat  g^{(0)}, \hat g^{(2)}$  as the AdS$_3$ one \cite{Skenderis:1999nb}, and  therefore it will have the same variation.

To include the effect of the  diffeomorphism that reinstates the radial gauge, we need to work at infinitesimal level above a background that has $A_\eta =0$. It is not hard to show that the first order contribution of $\xi^i$ to both the bulk and the boundary action vanishes.  Higher order contributions in $\xi^i$ will generically be non-zero; nevertheless, since $\xi^i$ is proportional to  $\eta = - \ell \log z$, the logarithm of the usual radial coordinate, the corrections to the on-shell action due to these terms will be purely logarithmic in  perturbation theory and can be renormalized separately from the power law divergences. 

 It would be interesting to understand  how the canonical momenta transform under the change of generalized coordinates from $(\g, A)$  to $\hat \g$. For this, we need to study the variation of the on-shell action. Since the computation becomes rather cumbersome in the general case, in the following we will only study  two simpler ones: the constant non-linear case
% where both metrics are in radial gauge and no diffeomorphism is needed, 
and that of a  general linearized perturbation around a given black string background.

%To compute the quadratic contribution, one needs the second order solution for $\xi^i$, which is rather cumbersome to obtain; alternatively, one can compute the \emph{variation} of the on-shell action, which receives contributions quadratic in $\xi^i$ but only requires knowing the linearized solution. 

\bigskip

\noindent \emph{$\diamond$ The constant case}

\medskip

\noindent In the constant (i.e. $x^i$-independent) case,  both $g$ and $\hat g$ are in radial gauge, since

\be
A_\eta = - \frac{\ell}{4} \hat \e^{ij} F_{ij} =0\;\;\;\;\;\; \Rightarrow \;\;\;\;\; \;
N_i =0 \;, \;\;\; N = \sqrt{1+\mu^2}
\ee
Therefore, we do not need to worry about the position of the radial cutoff surface. Setting $16 \pi G_3 =1$ for now, the variation of the on-shell action reads
\bea
\d S_{on-shell} & = & -  \int d^2 x \sqrt{\gamma} \left(  (K^{ij} -K \g^{ij} )\d \g_{ij}  + \frac{2}{\ell} \e^{ij} A_i \d A_j \right) = \label{dsonsh} \\ 
&  =&  -  \int d^2 x \sqrt{\hat \gamma} \left[ N^3 (K^{ij} -K \g^{ij}) \d \hat \g_{ij}  +2 \left( \frac{1}{\ell} \hat \e^{ij}- N^3 (K^{ij} -K \g^{ij})\right) A_i  \d A_j \right]\non
\eea
The terms proportional to $\d A_i$ in the above variation  are non-local in the  boundary data $\hat \g$, and  they
%The above variation does contain terms that are non-local in the chosen boundary data $\hat \g$, namely $\d A_i$.
% Thus, if the proposed boundary data are correct, these terms
 need to cancel from the final answer, for consistency.
%The inverse metric and the determinant read
%\be\g^{ij} = \frac{1}{1+\mu^2} \, \hat \g^{ij} + \hat A^i \hat A^j\;, \;\;\;\;\; A^i = \hat A^i \;, \;\;\;\;\; \det \g = (1+\mu^2) \det \hat \g \ee
%The extrinsic curvature is given by
%\be
%K_{ij} = - \frac{1}{2 N} \, \p_\eta \g_{ij} = - \frac{N}{2} (\p_\eta \hat \g_{ij} - A_i \p_\eta A_j - A_j \p_\eta A_i)
%\ee
%and, using the 
%\be\g^{ij} = \frac{1}{1+\mu^2} \hat \g^{ij} + \frac{1 + \mu^2}{1+ A_\eta^2 (1+\mu^2)^2} \hat A^i \hat A^j \;, \;\;\;\;\; \det \g = (1+\mu^2) (1+ (1+\mu^2)^2 A_\eta^2) \det \hat \g\ee
%The warped $AdS_3$ metric is then, simply
%\be ds^2 = (1+\mu^2)\, d\eta^2 + \g_{ij} (\eta) \, dx^i dx^j \;, \;\;\;\;\; \g_{ij} = (1+\mu^2)( \hat \g_{ij} - A_i A_j) \ee
%Other useful relations are
Using the simplified equations of motion for $A_i$
\be
\p_\eta A_i = \frac{2}{\ell} \, \hat \e_{ij} \hat A^j \;, \;\;\;\;\; A_i \hat A^i =\frac{\mu^2}{1+\mu^2}
\ee
one can show that the extrinsic curvature satisfies the following relations\footnote{The first relation follows %from the observation that $\hat A^i  \hat K_{ij} \hat A^j  =0$,  which implies that $K_{ij} \hat A^j$ and $\hat \e_{ij} \hat A^j $ are proportional. The constant of proportionality is fixed by computing the norm, and the remaining two relations in \eqref{KdotA} follow trivially from this one. This follows from 
the fact that both $\hat K_{ij} \hat A^j$ and $\hat \e_{ij} \hat A^j $  are normal to $A_i$ and their norm is constant. The norm can be computed  using the equation of motion for $\hat K_{ij}$: 
$
\p_\eta \hat K^{ij} = \hat R^{ij} -{}^{(2)} \hat R^{ij} + \hat K \hat K^{ij} + 2 \hat K^{ik} \hat K_k{}^j
$.
}

\be
\hat K_{ij} \hat A^j = - \frac{1}{\ell}\, \hat\e_{ij} \hat A^j \;, \;\;\;\;\;  
 K^{ij} = N^{-3} \hat K^{ij}\;, \;\;\;\;\;
K = N^{-1} \hat K\label{KdotA} 
\ee
%\be\d S = - \int d^2 x \sqrt{\gamma} \left( \g^{ij} \d K_{ij} + \d K + \frac{2}{\ell} \e^{ij} A_i \d A_j \right)\ee
%Adding the usual Gibbons-Hawking term, we now have
Using the above equations, we can show that

\bea
\d S & = &  -  \int d^2 x \sqrt{\hat \gamma} \left[  (\hat K^{ij} - \hat K \hat \g^{ij} - N^2 \hat K \hat A^i \hat A^j ) \d \hat \g_{ij} + 2 N^2 \hat K \hat A^j \d A_j \right] \non\\
& = & -  \int d^2 x \sqrt{\hat \gamma} \left[  (\hat K^{ij} - \hat K \hat \g^{ij})\, \d \hat \g_{ij} + N^2 \hat K \, \d (\hat A^i A_i) \right]
\eea
The last term vanishes by virtue of the fact that $\hat A^2$ is a constant. Thus, the T-mode part of the variation of the action is only proportional to $\d \hat \g_{ij}$, and the terms non-local in $\d \hat \g_{ij}$ have canceled away completely. In showing this, we only needed to use the equations of motion for the quantities multiplying the variations and did not add any counterterms. From the point of view of the canonical variables, we simply made the replacement

\be
\g_{ij},\; A_i \;\; \longrightarrow \;\; \hat \g_{ij} = N^{-2} \g_{ij} + A_i A_j , \; A_i
\ee
As a consequence, the canonical momenta changed to

\be
\pi^{ij} \r \pi^{ij} = N^{-2} \hat \pi^{ij} - \sqrt{\hat \g}\, \hat K\, \frac{\delta \mathcal{C}}{\d \hat \g_{ij} } \;, \;\;\;\;\;\;\; \pi^i_A \r \pi^i_A - 2 N^2 \pi^{ij} A_j = - N^2 \sqrt{\hat \g}\, \hat K \, \frac{\delta \mathcal{C}}{\d A_i } 
\ee
where $\mathcal{C} = \hat A^i A_i - \mu^2/ (1+\mu^2) =0$ is the constraint that the canonical variables satisfy and the canonical momenta can be read off from \eqref{dsonsh}. Adding the vanishing quantity $N^2  \hat K \,\mathcal{C} \,\sqrt{\hat \g} $ to the action, we find that the momentum conjugate to $\hat \g_{ij}$ becomes $\hat \pi^{ij}$, where

\be
\hat \pi^{ij} = - \sqrt{\hat \g} \, (\hat K^{ij} - \hat K \hat \g^{ij})
\ee
while the canonical momentum multiplying $\d A_i$ vanishes.  It would be interesting to check whether the above transformations correspond indeed to a canonical transformation in the constrained theory. To obtain the final renormalized momentum, one needs to add, as in AdS$_3$, a boundary cosmological constant counterterm, which renders the on-shell action finite. It also shifts $\hat \pi^{ij} \r \hat \pi^{ij} - \ell^{-1} \hat \g^{ij} \sqrt{\hat \g}$, thus making the map between the phase space variables $\hat \g_{ij}, \hat \pi^{ij} $ and the asymptotic constants $\hat g^{(0)}_{ij}, \hat g^{(2)\, ij}$ asymptotically diagonal \cite{papadimitriou}.

% Thus, using $K \sqrt{\g} = \hat K \sqrt{\hat \g}$,  we conclude that the variation of the on-shell action (without the Gibbons-Hawking term associated to $\g$ and having solved the constraint for $A_i$ in terms of $\hat \g$) needs the addition of a $\hat K$ Gibbons-Hawking term in order for the variational principle to be rendered well-defined, at least in the constant case. \emph{Out last sentence?}

\bigskip
\noindent \emph{$\diamond$ The non-constant case}
\medskip

\noindent We now consider linearized perturbations of the black string background \eqref{bhmet}. Since
 $A_{\eta} $ no longer vanishes, we need to take into account the effect of the diffeomorphism $\xi^i$ given by \eqref{formxi}.
%
% Even though we work at linearized order in $\xi_i$, $A_\eta$ and the $x^i$-derivatives of the various fields, the variation of the on-shell action can be computed at second order. 
%
%, and thus $\hat \g_{ij}$ and $\g_{ij}$ can no longer be simultaneously in radial gauge. Moreover, the linearized symplectic form satisfies $\p_{\eta} \Om =0$ only with respect to the Schr\"{o}dinger radial coordinate, and not the auxiliary AdS one. Thus, in order to construct the T-modes we need to start with the FG expansion in AdS with radial coordinate $\eta$, construct the warped AdS metric via \eqref{defhatg}, and then put the final metric in radial gauge. 
%
%We will perform the computation in two steps. We will first compute the variation of the on-shell action with respect to the metric $\g_{ij}$, and then separately evaluate the contribution of the diffeomorphism. Even though we work at linearized order in $\xi_i$, this contribution can be computed at second order if we evaluate the variation of the on-shell action  about a fixed background, rather than the action itself. We will choose the fixed background to be the warped black string solution \eqref{bhmet}. It all quantities that we compute, we will only keep the terms linear in $x^i$ derivatives, such as $A_\eta$, but not $A_\eta^2$, etc. 
%
After applying the diffeomorphism, the various quantities that enter the on-shell action, now denoted with a tilde, read
\be
\tilde \pi^{ij} = \pi^{ij} - \p_k \xi^i \pi^{kj} - \p_k \xi^j  \pi^{ik} + \xi^k \p_k \pi^{ij} \;, \;\;\;\;\; \tilde \g_{ij} = \g_{ij} + D_i \xi_j + D_j \xi_i
\ee

\be
\tilde A_i = A_i  + \p_i (\xi^k A_k) + \xi^k F_{ki}\;, \;\;\;\;\; \tilde A_\eta = A_\eta
\ee
Note that the last term in $\tilde \pi^{ij}$ and $\tilde A^i$ is  second order in derivatives and can be dropped.  The variation of the on-shell action is

\be
\d S  =  -  \int d^2 x \sqrt{\gamma} \left(  (\tilde K^{ij} - \tilde K \tilde \g^{ij} )\d \tilde \g_{ij}  + \frac{2}{\ell} \tilde \e^{ij} \tilde A_i \d \tilde A_j \right) = \d_0 S + \d_1 S + \d_2 S\ee
where $\d_0 S$ represents the action variation   before including the diffeomorphism, $\d_1 S$ is linear in $\xi^i$, while $\d_2 S$ is quadratic.
Of them, only  

\be
\d_0 S = -  \int d^2 x \sqrt{\gamma} \left(  (K^{ij} -K \g^{ij} )\d \g_{ij}  + \frac{2}{\ell} \e^{ij} A_i \d A_j \right)
\ee
contributes to the finite piece of the action. The remaining contributions only have logarithmic divergences and can be renormalized separately\footnote{Using extensively the fact that we have a linearized solution around an $x^i$-independent background, it is possible to show that $\d_2 S = 2 \eta^2 \mu^2 N^2 \int d^2 x \sqrt{\g} \, \pi^{ik} \p_i A_\eta \d \p_k A_\eta $ and thus can be canceled by a counterterm proportional to $(\log z)^2$ that is linear in the extrinsic curvature. We could not easily bring $\d_1 S$ to any nice form, but it is very possible that counterterms linear the holographic momenta will also be necessary.}. Noting that, at linearized level 

\be
\hat K_{ij} \hat A^j + \p_i A_\eta = - \frac{1}{\ell} \, \hat \e_{ij} \hat A^j
\ee
we can show that the expression for the on-shell action can be  simplified to

\bea
\d_0 S & = &  -  \int d^2 x \sqrt{\hat \gamma} \left[  (\hat K^{ij} - \hat K \hat \g^{ij}) \d \hat \g_{ij} + N^2 \hat K \d (\hat A^i A_i)+ 2 N^2 \hat D_k \hat A^k \d A_\eta \right] - \non \\
&& \hspace{3cm} -  2 N^2 \int d^2 x \,  \d (\hat A^k \p_k A_\eta \sqrt{\hat \g})
 \eea
Since  $A_\mu$  is divergence-free, we have

\be
 \hat D_k \hat A^k = \hat K A_\eta - \p_\eta A_\eta
\ee
Plugging in, we find

\bea
\d_0 S & = &  -  \int d^2 x \sqrt{\hat \gamma} \left[  (\hat K^{ij} - \hat K \hat \g^{ij}) \d \hat \g_{ij} + N^2 \hat K \d (\hat A^2)- 2 N^2 \p_\eta A_\eta \d A_\eta \right] - \non \\
&& \hspace{3cm} -  2 N^2 \int d^2 x \,  \d (\hat A^k \p_k A_\eta \sqrt{\hat \g})
 \eea
The first term is identical to the variation of the on-shell action in AdS$_3$, the second is zero upon imposing 
 the constraint, the third  vanishes identically in perturbation theory because $\p_\eta A_\eta =0$, and the forth agrees precisely with the linearization of \eqref{relK}. To render the action finite, we need to add the counterterms

\be
S_{ct} =   2 N^2 \int d^2 x \,  \hat A^k \p_k A_\eta \sqrt{\hat \g} - \frac{2}{\ell} \int d^2 x \,  \sqrt{\hat \g}
\ee
It would be interesting to understand whether the addition of the above counterterms  induces a canonical transformation in the constrained theory that describes the T-modes.

% Next, we need to add in the contribution of the diffeomorphism $\xi^i$. These read
% 
% \be
% \d_1 S = - \int d^2 x \sqrt{\g} \left[(\nabla_k \xi^k \pi^{ij} + \Delta \pi^{ij}) \d \g_{ij} + \pi^{ij} \d \Delta \g_{ij} + \frac{2}{\ell} \e^{ij} (\Delta A_i \d A_j + A_i \d \Delta A_j) \right]
% \ee
% 
% \be
% \d_2 S = - \int d^2 x \sqrt{\g} \left[ (\nabla_k \xi^k \pi^{ij} + \Delta \pi^{ij}) \d \Delta \g_{ij} + \frac{2}{\ell} \e^{ij} \Delta A_i \d \Delta A_j \right]
% \ee
% To simplify calculations, we plug in $\xi_i =  \eta\, N^2 A_i A_\eta$ and find
% 
% \bea
% \d_1 S &=& - \eta\int d^2 x \sqrt{\g} \left[  N^2  (A_0^k \pi_0^{ij} - A^i_0 \pi^{jk}_0 -A^j_0 \pi^{ik}_0) \p_k A_\eta \d \g_{ij} + \frac{2 \mu^2}{\ell} \e^{ij} \p_i A_\eta \d A_j \right.\non\\
% && \left. - 2 (N^2 \p_i \pi^{ij} A_j^0 + \frac{\mu^2}{\ell} \e^{ij} \p_j A_i) \d A_\eta\right]
% \eea
% 
% \bea
% \d_2 S &= &\eta^2 N^2 \int d^2 x \sqrt{\g} \left[ 2 N^2 A_j (A^k \pi^{ij} - A^i \pi^{jk}-A^j \pi^{ik}) \p_i \p_k A_\eta \d A_\eta + 2 \mu^2 \e^{ij} \p_i \p_j A_\eta \d A_\eta \right] \non\\
% &=& 2 \eta^2 \mu^2 N^2 \int d^2 x \sqrt{\g} \pi^{ik}_0 \p_i A_\eta \d \p_k A_\eta
% \eea
% which is the linearization of a counterterm linear in $\hat K$.

\subsection{The holographic stress tensor}

Since the renormalized on-shell action in the warped black string backgrounds equals the one in the auxiliary AdS$_3$ and the boundary data are the same,
 it follows that the holographic stress tensor is the same as the AdS$_3$ one

\be
\langle T_{ij} \rangle  = - \frac{2}{\sqrt{\hat g^{(0)}}} \frac{\d S_{ren} [\hat g^{(0)}]}{\d \hat g^{(0)\, ij}} = \frac{\ell}{8 \pi G_3}
(\hat g^{(2)}_{ij} - \hat g^{(0)}_{ij} \hat g^{(2) k}{}_k) \label{holotij}
\ee
Using the asymptotic relations \eqref{constrg} between $\hat g^{(2)}$ and $\hat g^{(0)}$, we find the holographic Ward identities
\be
\hat \nabla_i \hat T^{ij} = 0 \;, \;\;\;\;\; \hat T^i_i = \frac{\ell}{16 \pi G_3} \hat R[\hat g^{(0)}] \label{holoward}
\ee
The fact that the stress tensor is both symmetric and conserved is indicative of an underlying relativistic structure  of this sector of the theory, which is consistent with the existence of both left-moving and right-moving Virasoro extensions of the asymptotic symmetry group \cite{stromwei,trunc}. The conformal anomaly read off from the above equation

\be
c = \frac{3\ell}{2 G_3}
\ee
agrees with the result \eqref{cc} obtained using  asymptotic symmetry group analyses. We can also compute 
the expectation value of the holographic stress tensor on the black string backgrounds 
\be
\langle T_{++} \rangle = \frac{\ell}{8 \pi G_3} T_+^2 \;, \;\;\;\;\; \langle T_{--} \rangle = \frac{\ell}{8 \pi G_3} T_-^2
\ee
The energy and momentum of the string are given by

\be
Q_\xi = \int dx \, \hat n^i \, \langle T_{ij} \rangle\, \xi^j \sqrt{\hat \s} \;, \;\;\;\;\; \xi^i = \p_\tau, \p_x \label{cchar} 
\ee
where the coordinates $x,\tau$ are defined via
\be
x^\pm = x \pm \tau
\ee
$\hat n^i$ is the unit normal to the surface of constant $\tau$ at the boundary (using the metric $\hat \g$) and $\hat \s$ is the induced metric on this surface. It is not hard to show that $Q_\tau$ and $Q_x$ per unit $x$-length equal precisely the energy and momentum per unit length of the string \eqref{epmp}, which were computed in \cite{trunc} using covariant methods. 

It is quite remarkable that a simple, linear expression such as \eqref{cchar} can capture the conserved charges, which in the covariant formalism receive non-linear contributions that are  strongly background-dependent.
% This  simplicity may be related to the fact that
Note that 
the conserved charges \eqref{cchar} effectively ``live'' in the auxiliary AdS$_3$ spacetime used for the construction of the T-modes, whose definition depends on the value of $T_+$. It is plausible that by formulating the problem in terms of the auxiliary AdS$_3$ spacetime, the non-linear charges in warped AdS$_3$ that one always encounters in the covariant formalism are traded for a set
of linear charges\footnote{We thank Tom Hartman for making this point. }, but with a state-dependent dictionary \eqref{defhatg}.

%It would be very interesting to better understand the relationship between these two different methods of computing conserved quantities.
% It would be much more desirable to find conserved charges that are constructed from the asymptotic metric in the warped AdS$_3$ spacetime itself. We leave this interesting problem to future investigations. 

\subsection{General warped black strings \label{genbsb}}

\noindent It is not hard to extend these results to the general four-parameter family of warped black strings in type IIB supergravity with RR three-form flux found in \cite{trunc}. These six-dimensional black string solutions depend as before on two parameters $T_\pm$ related to their Hawking temperature and horizon velocity via \eqref{tempom}. In addition, they have  two  warping parameters $\l_{1,2}$, which correspond to two distinct massive vector fields in spacetime, and consequently  two different $(1,2)$  operators  that define the boundary theory. 

The solutions have a somewhat complicated dependence on the two parameters $\l_{1,2}$, so we will not reproduce them here\footnote{They can be found in the equations (2.19) - (2.24) in \cite{trunc}, whose notation we are using. The length $\hat \ell$ is defined in (3.23) of the same reference.}. Nevertheless, the energy and momentum per unit length of any given black  string take the following simple form \cite{trunc}

\be
E \pm P = \frac{ \pi \hat \ell^4}{2  G_6} T_\pm^2 
\ee
where $\hat \ell^4$  equals the product of the electric and magnetic charges of the  string. The asymptotic symmetry group computations of \cite{trunc} yield a Virasoro algebra with central charge

\be
c = \frac{3\pi^2 \hat \ell^4}{G_6}
\ee 
We would like to reproduce these values from a holographic stress tensor computation. 

\bigskip

The six-dimensional black string solutions of \cite{trunc} cannot in general be captured by a three-dimensional consistent truncation of type IIB string theory; nevertheless, if we only consider T-modes about a given black string background with temperature $T_+$, then such a truncation is possible.  Upon the dimensional reduction, the three-dimensional Einstein metric can be written as

\be
ds_3^2 =(1+\mu^2) \, ds^2_{BTZ} - \mu^2 \hat \ell^2 \s_0^2  \;, \;\;\;\;\; \mu = \tilde \l T_+
\ee
where the BTZ radius is $\hat \ell$ and the relationship between the six- and three-dimensional Newton constants is $G_6 = 2 \pi^2 \hat \ell^3 G_3$. The quantity $\tilde \lambda$ is defined in \cite{trunc} and depends on $\l_1, \l_2 $ and $T_+$ in a rather complicated manner. The two massive vector fields that support the solution are both proportional to an auxiliary massive vector field satisfying \eqref{eqA} and given by
\be
A_\mu = \frac{\mu \hat \ell}{\sqrt{1+\mu^2}} \s_0 
\ee 
The $SL(2,\mathbb{R})_L \times U(1)_R$-invariant one-form $\s_0$ needs to be written in black hole coordinates, using \eqref{cootransf}. The reason that $\hat \ell$, rather than any other length scale, is singled out is because the on-shell six-dimensional action is proportional to $\hat \ell^4$ and thus it is the natural  length parameter to keep fixed in the reduction.

We now define the T-modes as before in \eqref{defhatg} - \eqref{eqA}, with the only change that we replace $\ell$ by $\hat \ell$. Using the explicit solutions of \cite{trunc}, %which are too cumbersome to write down here, 
we can show that the on-shell bulk action, when evaluated on any T-mode solution, equals

\be
S_{bulk} = - \frac{1}{16 \pi G_3} \int d^3 x \sqrt{\hat g}\, \frac{4}{\hat \ell^2  }
\ee 
The boundary action is the usual Gibbons-Hawking term and holographic renormalization on the truncated T-mode sector works identically as before. Thus, we obtain again a stress tensor of the form \eqref{holotij} that satisfies the holographic Ward identities \eqref{holoward}, with conformal anomaly
\be
c = \frac{3 \hat \ell }{2 G_3} = \frac{3 \pi^2 \hat \ell^4}{G_6}
\ee
and holographic expectation values

\be
\langle T_{++} \rangle = \frac{\hat \ell}{8 \pi G_3} T_+^2 = \half (E +P) \;, \;\;\;\;\; \langle T_{--} \rangle = \frac{\hat \ell}{8 \pi G_3} T_-^2 = \half (E-P)
\ee
again in perfect agreement with the covariant charge computations of \cite{trunc}.

\section{Discussion \label{disc}}

The goal of this article was to find the holographic interpretation of  the general warped black strings described in  \cite{trunc}, i.e. to spell out which operator sources and expectation values are turned on in these backgrounds. The holographic dictionary we have proposed is based upon  Papadimitriou's prescription \cite{papadimitriou} and a certain  reasonable assumption \eqref{deftemp} about the behaviour of  finite-temperature dipole CFTs. Our holographic computation of the conserved charges and anomalies agrees with previous results obtained using covariant methods \cite{trunc}.

The warped AdS$_3$ holographic dictionary   we have  proposed   has a number of curious features, such as the  state-dependent identification of the  holographic sources. It would be interesting to better understand this trait. Also, we need to better study the structure of the counterterms  necessary for  holographic renormalization.  For example, in AdS the counterterms need to be constructed from the induced fields at the boundary and their derivatives \cite{Skenderis:2002wp}. For spacetimes corresponding to irrelevant deformations of AdS, \cite{vanRees:2011fr,vanRees:2011ir,vanRees:2012cw} argued that the counterterms can also include the conjugate momenta (i.e. the extr‌insic curvature), but only at quadratic order or higher.  It would be interesting to understand the general rules for constructing counterterms in warped AdS$_3$, renormalized according to the prescription of \cite{papadimitriou}. 

Another interesting exercise would be to check whether holographic renormalization in warped AdS$_3$  - with holographic sources chosen \`a la Papadimitriou  -  corresponds  indeed to a canonical transformation, thus following the pattern described in  \cite{papadimitriou}. Should this be the case, it would be a non-trivial 
check that the prescription is self-consistent even in a case where its predictions are different from those obtained using other methods \cite{Ross:2009ar,vanRees:2012cw}.  For better understanding this issue, it may be useful to also include the X-modes in the analysis, which can be easily done, at least at linearized level. 

Although we have been able to find agreement between the energy and momentum density of the black strings computed using the holographic versus the covariant method, the relationship between the two  formalisms is far from clear. For example, the holographic method uses  
% our construction of  conserved charges in a general  warped AdS$_3$ background is far from satisfactory. The reason is that it uses
 the auxiliary AdS$_3$ metric $\hat \g$ instead of the induced metric, $\g$, at the boundary, thus completely obscuring
  the conservation of the charges from the point of view of warped AdS$_3$. On the other hand, the holographic method yields charges that are linear in the stress tensor and take a very simple form in the auxiliary AdS$_3$ spacetime, to be compared with the very non-linear expressions for the charges in the covariant formalism. It would thus be very interesting to understand the precise relationship between the two formalisms and use it to  understand e.g. the Kerr/CFT correspondence - where the covariant methods of \cite{Barnich:2001jy} have been particularly successful - from a holographic point of view. 

Finally, one should ascertain whether Papadimitriou's prescription, which has led to the above rather interesting and intriguing results, %(stress tensor both symmetric and conserved, state-dependent dictionary) 
is also the correct one. This question can only be answered by understanding the relationship between the spacetime fields and the field theory gauge-invariant operators. The latter may be possible to construct using the prescription of \cite{Gross:2000ba} for non-commutative field theories, and the couplings to the spacetime metric may be found using the methods of \cite{Okawa:2000sh,Liu:2001ps}.
% Given the similarity between dipole theories, which  are the field theory duals to certain Schr\"{o}dinger spacetimes, and non-commutative theories, it is possible that the works of \cite{} are useful in answering this question. 
Should the answer be affirmative, it would have implications not only for warped AdS$_3$ holography and the Kerr/CFT correspondence, but it would also give us confidence that Papadimitriou's prescription is the correct general one. %prescription to apply also in other spacetimes, such as for example flat space.

\bigskip

\noindent \textbf{Acknowledgements}

\medskip

\noindent The author is grateful to Tom Hartman, Juan Maldacena, Ioannis Papadimitriou, Andrew Strominger and Marika Taylor for interesting conversations, and would like to especially thank Tom Hartman for useful comments on the draft. This work was supported in part by the DOE grant DE-SC0007901.

\end{document}